\RequirePackage{fix-cm}
\documentclass[smallextended]{svjour3}       
\smartqed  
\usepackage{graphicx}
\usepackage{mathptmx}      
\usepackage{latexsym}
\usepackage{amsmath}
\usepackage{amssymb}
\usepackage{natbib}
\RequirePackage[colorlinks,citecolor=blue,urlcolor=blue]{hyperref}

\newcommand{\Real}{\mathbb R}
\newcommand{\pr}{{\rm Pr}}

\newcommand{\calD}{{\mathcal{D}}}
\newcommand{\calE}{{\mathcal{E}}}

\newcommand{\calP}{{\mathcal{P}}}
\newcommand{\calS}{{\mathcal{S}}}
\newcommand{\dth}{{$d{\rm th}$\;}}

\begin{document}

\title{Likelihood estimators for multivariate extremes
}


\author{Rapha\"el Huser \and Anthony C. Davison \and\\
Marc G. Genton}


\institute{R. Huser \at
              CEMSE Division, King Abdullah University of Science and Technology, Thuwal 23955-6900 (SA)\\
              Tel.: +966-12-8080682\\
              \email{raphael.huser@kaust.edu.sa} 
            \and
           A.~C. Davison \at
              EPFL SB MATHAA STAT, Station 8, B\^atiment MA, 1015 Lausanne (CH)\\
	      Tel.: +41-21-6935502\\
	      \email{anthony.davison@epfl.ch}   
           \and
           M.~G. Genton \at
           CEMSE Division, King Abdullah University of Science and Technology, Thuwal 23955-6900 (SA) \\
	  Tel.: +966-12-8080244\\
           \email{marc.genton@kaust.edu.sa}
}

\date{Received: date / Accepted: date}

\maketitle

\begin{abstract} The main approach to inference for multivariate extremes consists in approximating the joint upper tail of the observations by a parametric family arising in the limit for extreme events. The latter may be expressed in terms of componentwise maxima, high threshold exceedances or point processes, yielding different but related asymptotic characterizations and estimators. The present paper clarifies the connections between the main likelihood estimators, and assesses their practical performance. We investigate their ability to estimate the extremal dependence structure and to predict future extremes, using exact calculations and simulation, in the case of the logistic model.

\keywords{Asymptotic relative efficiency \and Censored likelihood \and Logistic model \and Multivariate extremes \and Pairwise likelihood \and Point process approach}
\end{abstract}

\section{Introduction}
\label{IntroSection}
Under mild conditions, multivariate extreme-value distributions provide models suitable for the stochastic fluctuations of componentwise maxima. The limiting distribution of linearly renormalized componentwise maxima of independent and identically distributed (i.i.d.) random vectors, provided it exists and is non-degenerate, is necessarily a multivariate extreme-value distribution \citep[][Chapter 5]{Resnick:1987}. Although these distributions have parametric generalized extreme-value (GEV) margins, their dependence structure is non-parametric. A standard approach to inference for multivariate extremes consists in approximating the distribution of componentwise \emph{finite}-block maxima by flexible parametric \emph{asymptotic} submodels, proposed among others by \citet{Gumbel:1961}, \citet{Tawn:1988b}, \citet{Husler.Reiss:1989}, \citet{Coles.Tawn:1991}, \citet{Joe.etal:1992}, \citet{Demarta.McNeil:2005} and \citet{Segers:2012}. In order to increase computational and statistical efficiency, \citet{Stephenson.Tawn:2005} proposed a refined approach which uses the extra information of occurrence times of extreme events, a biased-corrected version of which is proposed by \citet{Wadsworth:2015}. In high dimensions, composite likelihoods \citep{Lindsay:1988,Varin.etal:2011} may also reduce the computational burden, while retaining fairly high efficiency. Non-parametric estimation procedures have also been considered \citep[see, e.g.,][]{Pickands:1981,Deheuvels.deOliveira:1989,Deheuvels:1991,Smith.etal:1990,Caperaa.etal:1997,Hall.Tajvidi:2000,Boldi.Davison:2007}, but in the present paper we focus on parametric approaches.

An alternative approach, the point process characterization of extremes \citep{Coles.Tawn:1991}, enables efficient inference by incorporating additional data, which are lower than block maxima in a sense to be made precise below, but sufficiently large to provide useful information about extremal characteristics. Loosely speaking, in practice this approach consists of fitting a non-homogeneous Poisson process to high threshold exceedances. In the univariate framework, this is essentially the same as fitting a generalized Pareto distribution (GPD) to exceedances \citep{Davison.Smith:1990,Smith:1989}, and it extends to higher dimensions through the multivariate GPD \citep{Falk.Reiss:2005,Rootzen.Tajvidi:2006,Buishand.etal:2008}. In the multivariate framework, the notion of exceeding a given threshold may be interpreted in various ways, thereby yielding different threshold-based estimators \citep[see, e.g.,][]{Coles.Tawn:1991,Resnick:1987,Beirlant.etal:2004,Fougeres:2004}. Alternatively, noticing that the dependence structure of high threshold exceedances is essentially the same as that of componentwise maxima, \citet{Ledford.Tawn:1996} and \citet{Smith.etal:1997} proposed a censored likelihood, decreasing the contribution of points that are ``not extreme enough''; see also \citet{Bortot.etal:2000}, \citet{Thibaud.etal:2013} and \citet{Huser.Davison:2014a}.

Although all the aforementioned estimators are closely linked to each other, in the sense that they may be viewed as stemming from the same asymptotic result, they nevertheless have different properties in practice. In general, block maximum approaches may be expected to be relatively unbiased but rather variable, whereas threshold-based approaches are commonly thought to be more efficient, but more biased. However, as far as we know, no quantitative study of their performance has yet been performed, though \citet{Zheng.etal:2014} is a related recent contribution.

The goal of the present paper is to clarify the connections between the main likelihood estimators for multivariate extremes, and to provide a quantitative assessment of their performance. We focus on the estimation of the dependence structure, rather than the marginal distributions, and the logistic extreme-value model is considered for its simplicity and tractability in high dimensions. In Section \S\ref{MultivariateExtremesSection}, an overview of classical results in multivariate extreme-value theory is given, and the symmetric and asymmetric logistic families are presented. In \S\ref{InferenceSection}, likelihood estimators are described, and in \S\ref{PerformanceAssessmentSection}, their performance is assessed using analytical calculations and simulation based on the logistic model. Finally, \S\ref{DiscussionSection} contains some discussion.

\paragraph{Vector notation.} Throughout the paper, bold symbols denote $D$-dimensional random or deterministic real vectors. For example, $\vec{Y}=(Y_1,\ldots,Y_D)^T$, $\vec{a}_n=(a_{n,1},\ldots,a_{n,D})^T$, $\vec{0}$ is a $D$-dimensional vector of zeros, $\pmb{\infty}$ is a vector of infinities, etc. All vector operations are componentwise: $\vec{y}\leq\vec{u}$ means $y_d\leq u_d$ for all $d=1,\ldots,D$, $\vec{a}\vec{y}$ is a vector with \dth component $a_dy_d$, $\max_{i=1}^n \vec{Y}_i=(\max_{i=1}^n Y_{i,1},\ldots, \max_{i=1}^n Y_{i,D})^T$, etc. Furthermore, $\vec{y}\nleq\vec{u}$ indicates that there exists at least one $d=1,\ldots,D$ such that $y_d>u_d$. If a comparison or an operation is done between a vector and a scalar, it holds for each component of the vector: $a\vec{y}$ is a vector with components $ay_d$, etc. When sets are involved, $[\vec{a},\vec{b})$ is the product set $[a_1,b_1)\times\cdots\times[a_D,b_D)$.

\section{Multivariate extremes}
\label{MultivariateExtremesSection}

\subsection{Asymptotic theory and upper-tail approximations}
\label{AsymptoticTheorySection}
This section summarizes some of the main results of multivariate extreme-value theory. More detailed surveys may be found in \citet{Resnick:1987}, \citet{Coles:2001}, \citet{Fougeres:2004}, \citet{Beirlant.etal:2004}, \citet{Segers:2012}, \citet{Davison.Huser:2015} and the references therein. 

Let $\vec{Y}$ denote a $D$-dimensional random vector with joint distribution function $F(\vec{y})$ and margins $F_d(y)$ $(d=1,\ldots,D)$. Moreover, let $\vec{Y}_i$ $(i=1,2,\ldots)$ denote a sequence of i.i.d. replicates of $\vec{Y}$ and consider the vector of componentwise maxima $\vec{M}_n=\max_{i=1}^n\vec{Y}_i$. A key goal of multivariate extreme-value theory is to characterize the family of asymptotic distributions that arise as limits for $\vec{M}_n$, when suitably renormalized by location and scale sequences. Hence, assume that sequences $\vec{a}_n\in\Real_+^D$ and $\vec{b}_n\in\Real^D$ may be found such that as $n\to\infty$ the sequence of renormalized maxima $\vec{a}_n^{-1}(\vec{M}_n-\vec{b}_n)$ converges in distribution to a random vector $\vec{Z}$ with joint distribution $G(\vec{z})$ and non-degenerate margins $G_d(z)$ $(d=1,\ldots,D)$. If such sequences exist, we say that $\vec{Y}$ is in the max-domain of attraction of $\vec{Z}$, and the limiting distribution function may be expressed as
\begin{equation}\label{ResultMaxima}
G(\vec{z})=\exp\left[-V\{t(\vec{z})\}\right]
\end{equation}
and is called a multivariate extreme-value distribution. The function $V$ on the right-hand side of (\ref{ResultMaxima}), called the exponent measure, is homogeneous of order $-1$, i.e., $V(s\vec{z}^\star)=s^{-1}V(\vec{z}^\star)$ for any $s>0$ and any $\vec{z}^\star>\vec{0}$, and satisfies the marginal constraints $V(\infty,\ldots,\infty,z^\star,\infty,\ldots,\infty)=1/z^\star$ for any permutation of the $D$ arguments. The function $t(\vec{z}):\Real^D\to\Real^D_+$ in (\ref{ResultMaxima}) is a marginal transformation which, provided $\vec{a}_n$ and $\vec{b}_n$ are suitably chosen, maps the vector $\vec{z}$ to $\{t_1(z_1),\ldots,t_D(z_D)\}^T$, where
\begin{equation}\label{MarginalTransformation}
t_d(z)=\left(1+\xi_d z\right)_+^{1/\xi_d},\qquad d=1,\ldots,D,
\end{equation}
with $a_+=\max(0,a)$, and $\xi_d\in\Real$. This implies that for large $n$, the marginal distributions of $\vec{M}_n$ are approximately GEV with location parameter $b_{n,d}$, scale parameter $a_{n,d}$ and shape parameter $\xi_d$, i.e., writing $t_{n,d}(z)=t_d\{(z-b_{n,d})/a_{n,d}\}$,
\begin{eqnarray}
\pr(M_{n,d}\leq z)&\approx& G_d\left({z-b_{n,d}\over a_{n,d}}\right)\;\;=\;\;\exp\left\{-1/t_{n,d}(z)\right\}\nonumber\\
&=&\exp\left\{-\left(1+\xi_d{z-b_{n,d}\over a_{n,d}}\right)_+^{-1/\xi_d}\right\},\qquad d=1,\ldots,D.\label{GEV}
\end{eqnarray}
Since the variates $Z_d^\star=t_d(Z_d)$ all have unit Fr\'echet distributions, meaning that $\pr(Z_d^\star\leq z^\star)=\exp(-1/z^\star)$, $z^\star>0$, the functions in (\ref{MarginalTransformation}) may be used to transform the data to a common scale, thereby enabling separate treatment of the margins and the dependence structure. A key point for the proof of (\ref{ResultMaxima}) is that the class of multivariate extreme-value distributions coincides exactly with max-stable distributions $G(\vec{z})$ with non-degenerate margins, meaning that there exist $\vec{a}_k\in\Real_+^D$ and $\vec{b}_k\in\Real^D$ such that
\begin{equation*}\label{MaxStability}
G^k(\vec{a}_k \vec{z} + \vec{b}_k) = G(\vec{z}),\qquad k=1,2,\ldots.
\end{equation*}
Hence, $G(\vec{z})$ is also max-infinitely divisible: it can be viewed as the distribution of the maximum of $k$ i.i.d. random variates for any positive integer $k$. Therefore, according to \citet{Balkema.Resnick:1977} and \citet[][p.255]{Beirlant.etal:2004}, there must exist a unique measure $\nu$ concentrated on $\Omega=[\vec{c},\pmb{\infty})\setminus\vec{c}$ for some $\vec{c}\in[-\pmb{\infty},\pmb{\infty})$, such that 
\begin{equation}\label{MaxInfiniteDivisibility}
G(\vec{z}) = \exp\{-\nu(A_{\vec{z}})\},\quad \vec{z}\in\Omega,
\end{equation}
where $A_{\vec{z}}$ denotes the complement of the set $[-\pmb{\infty},\vec{z}]$ in $[-\pmb{\infty},\pmb{\infty})$. Since the limiting marginal distributions are necessarily GEV \citep{Fisher.Tippett:1928}, the measure $\nu$, transformed using (\ref{MarginalTransformation}), yields a measure $\nu_t$ on $[\vec{0},\pmb{\infty})\setminus\{\vec{0}\}$ such that
\begin{equation}\label{ExponentMeasureFrechet}
\nu(A_{\vec{z}})=\nu_t\{A_{t(\vec{z})}\}=V\left\{t(\vec{z})\right\},
\end{equation}
thereby recovering (\ref{ResultMaxima}). Moreover, the homogeneity of $V$ is a direct consequence of the max-stability of $G(\vec{z})$.

Deeper insight may be obtained by considering extreme events from a point process perspective. Assuming that (\ref{ResultMaxima}) holds, consider the point process
\begin{equation}\label{PointProcess}
P_n = \left\{{\vec{Y}_i-\vec{b}_n\over \vec{a}_n}: i=1,\ldots,n\right\}.
\end{equation}
According to \citet[][p.154]{Resnick:1987}, as $n\to\infty$, $P_n$ converges to a non-homogeneous Poisson process $P$ on $\Omega$ with mean measure $\nu$, as defined in (\ref{MaxInfiniteDivisibility}). Thanks to Equation (\ref{ExponentMeasureFrechet}), if the measure $\nu$ is absolutely continuous, the corresponding intensity measure is $\nu({\rm d}\vec{y})=-|J_t(\vec{y})|V_{1:D}\{t(\vec{y})\}{\rm d}\vec{y}$, where $V_{1:D}$ is the derivative of the function $V$ with respect to all arguments, and $J_t(\vec{y})$ is the Jacobian associated to the transformation $t(\vec{y})$. By the Poisson property, one has that for any Borel set $B\subset\Omega$ with compact closure and zero mass on its boundary \citep[][p.280]{Beirlant.etal:2004},
\begin{equation}\label{PointProcess2}
\pr(P_n \subset \Omega\setminus B)\to\pr(P\subset \Omega\setminus B)=\exp\{-\nu(B)\},\quad n\to\infty,
\end{equation}
and by choosing $B=A_{\vec{z}}=[-\pmb{\infty},\pmb{\infty})\setminus[-\pmb{\infty},\vec{z}]$, for $\vec{z}>\vec{c}$, (\ref{PointProcess2}) combined with (\ref{ExponentMeasureFrechet}) yields (\ref{ResultMaxima}). Furthermore, $\vec{Y}$ is in the max-domain of attraction of $G(\vec{z})$ if and only if
\begin{equation}\label{ConvergenceMeasure}
\nu_n(B)=n\pr\left({\vec{Y}-\vec{b}_n\over\vec{a}_n}\in B\right)\to\nu(B),\quad n\to\infty,
\end{equation}
for any Borel set $B\subset\Omega$ defined in \eqref{PointProcess2}. As a result, replacing the convergence in (\ref{ConvergenceMeasure}) by equality for large $n$, letting $\vec{u}\in\Omega$ be a high threshold (typically of the form $\vec{u}=\vec{a}_n\vec{u}^\circ+\vec{b}_n$ for some fixed $\vec{u}^\circ$) and choosing $B=A_{\vec{y}}=[-\pmb{\infty},\pmb{\infty})\setminus[-\pmb{\infty},\vec{y}]$, one obtains the upper tail approximation
\begin{equation}\label{ApproximationTail}
F(\vec{y})\approx1-{1\over n}V\left\{t\left({\vec{y}-\vec{b}_n\over \vec{a}_n}\right)\right\}=1-V\{nt_n(\vec{y})\}\approx\exp[-V\{\tilde{t}_n(\vec{y})\}],\quad \vec{y}>\vec{u}.
\end{equation}
Here we have used the homogeneity of the exponent measure, (\ref{ExponentMeasureFrechet}), and a first order Taylor expansion of the exponential function, and $\tilde{t}_n(\vec{y})=nt_n(\vec{y})=\{nt_{n,1}(y_1),\ldots,nt_{n,d}(y_d)\}^T$ denotes the marginal transformation defined in \eqref{MarginalTransformation} and \eqref{GEV} with modified location and scale parameters; specifically, the $d$th location parameter is $b_{n,d}+a_{d,n}(n^{-\xi_d}-1)/\xi_d$, the $d$th scale parameter is $a_{d,n} n^{-\xi_d}$, but the shape parameter $\xi_d$ remains unchanged. Hence, whenever (\ref{ResultMaxima}) holds, the upper tail of the distribution of $\vec{Y}$ may be approximated by a multivariate extreme-value distribution with essentially the same dependence structure as maxima.

It is useful to represent a random variate $\vec{Z}$ distributed according to (\ref{ResultMaxima}) in terms of pseudo-polar coordinates,
\begin{equation*}\label{PseudoPolar}
R=\sum_{d=1}^D Z_d^\star=\sum_{d=1}^D t_d(Z_d),\quad \vec{W}={\vec{Z}^\star\over R}={t(\vec{Z})\over R}.
\end{equation*}
Here $R$ represents the radial part, i.e., the overall magnitude of $\vec{Z}$ on the unit Fr\'echet scale, and $\vec{W}$ denotes the vector of relative magnitudes of each component. One can show \citep[][p.258]{Beirlant.etal:2004} that the limiting intensity measure factorizes as
\begin{equation}\label{PoissonLimitingMeasure}
\nu({\rm d}\vec{y})=\nu({\rm d}r,{\rm d}\vec{w})=Dr^{-2}{\rm d}r\;H({\rm d}\vec{w}),
\end{equation}
where $H$ is a probability measure on the $(D-1)$-dimensional simplex $\calS_D=\{\vec{w}\in[0,1]^D:\sum_{d=1}^D w_d=1\}$, satisfying the mean constraints $\int_{\calS_D} w_d H({\rm d}\vec{w})=D^{-1}$ for $d=1,\ldots,D$. The measure $H$ is called the spectral measure, and if it is absolutely continuous, then its Radon--Nikodym derivative $h(\vec{w})$ is called the spectral density. Relation (\ref{PoissonLimitingMeasure}) implies that the angular and radial components are asymptotically independent. Furthermore, it follows from (\ref{ExponentMeasureFrechet}) and (\ref{PoissonLimitingMeasure}) that the exponent measure may be expressed as
\begin{equation}\label{ExponentMeasure}
V(\vec{z}^\star)=\nu\{A_{t^{-1}(\vec{z}^\star)}\}=\int_{\calS_D}\int_{\min(\vec{z}^\star/\vec{w})}^\infty D{{\rm d}r\over r^2} H({\rm d}\vec{w})= D\int_{\calS_D}\max\left({\vec{w}\over \vec{z}^\star}\right) H({\rm d}\vec{w}).
\end{equation}
Similarly, considering the extreme set $A^{\vec{r}_0}=\{\vec{z}\in\Omega:\sum_{d=1}^Dt_d(z_d)/r_{0,d}>1\}$, one has
\begin{equation}\label{ExponentMeasure2}
\nu(A^{{\vec{r}_0}})=\int_{\calS_D}\int_{\left\{\sum_{d=1}^D w_d/r_{0,d}\right\}^{-1}}^\infty D{{\rm d}r\over r^2} H({\rm d}\vec{w})= D\int_{\calS_D}\sum_{d=1}^D {w_d\over r_{0,d}}H({\rm d}\vec{w}) = \sum_{d=1}^D r_{0,d}^{-1},
\end{equation}
which, unlike (\ref{ExponentMeasure}), does not depend on $H$.

A consequence of the point process characterization is that the multivariate extension of the GPD is the limiting distribution for threshold exceedances. Specifically, assume that (\ref{ResultMaxima}) holds and let $\vec{u}^\circ\in\Omega$ denote some threshold vector on the renormalized scale. From (\ref{ExponentMeasureFrechet}) and (\ref{ConvergenceMeasure}), one can show that, as $n\to\infty$,
\begin{equation}\label{mGPD}
\pr\left({\vec{Y}-\vec{b}_n\over \vec{a}_n}\leq \vec{y}\;\bigg|\; {\vec{Y}-\vec{b}_n\over \vec{a}_n}\nleq \vec{u}^\circ\right)\to{V[\min\{t(\vec{y}),t(\vec{u}^\circ)\}]-V\{t(\vec{y})\}\over V\{t(\vec{u}^\circ)\}},
\end{equation}
the right-hand side of which may be rewritten using $G(\vec{y})=\exp\left[-V\{t(\vec{y})\}\right]$ as
\begin{equation}\label{mGPD2}
Q(\vec{y})={1\over -\log\{G(\vec{u}^\circ)\}}\log\left[{G(\vec{y})\over G\{\min(\vec{y},\vec{u}^\circ)\}}\right],\quad\vec{y}\nleq\vec{u}^\circ,
\end{equation}
known as a multivariate GPD with reference vector $\vec{u}^\circ$ \citep{Falk.Reiss:2001,Falk.Reiss:2002,Falk.Reiss:2003a,Falk.Reiss:2003b,Falk.Reiss:2005,Rootzen.Tajvidi:2006,Buishand.etal:2008}. If the density of $Q(\vec{y})$ exists, then it equals $q(\vec{y})=-|J_t(\vec{y})|V_{1:D}\{t(\vec{y})\}/V\{t(\vec{u}^\circ)\}$ $(\vec{y}\nleq\vec{u}^\circ)$, where $V_{1:D}(\vec{y})=\partial^D V(\vec{y})/\partial y_1\cdots\partial y_D$, and $J_t(\vec{y})$ is the Jacobian of the marginal transformation $t(\vec{y})$. It can be verified that if a random vector $\vec{Y}=(Y_1,\ldots,Y_D)^T$ is distributed according to $Q(\vec{y})$ in \eqref{mGPD2}, then the \dth conditional marginal distribution of exceedances may be expressed as
\begin{equation}\label{GPD}
\pr(Y_d\leq y\mid Y_d>u_d^\circ)=1-\left(1+\xi_d{y-u_d^\circ\over\tau_d}\right)_+^{-1/\xi_d},\quad y>u^\circ_d,
\end{equation}
where $\tau_d=1+\xi_du_d^\circ>0$; (\ref{GPD}) is a univariate GPD with location parameter $u^\circ_d$, scale parameter $\tau_d$ and shape parameter $\xi_d$. In addition, using the law of total probability, \eqref{mGPD} yields the following tail approximation, for large $n$ and large thresholds $\vec{u}$,
\begin{equation}\label{ApproximationTail2}
F(\vec{y})\approx1-V\{nt_n(\vec{y})\},\quad\vec{y}>\vec{u},
\end{equation}
which coincides with the middle approximation in (\ref{ApproximationTail}). This shows that multivariate extreme-value and multivariate GPD approximations to the upper tail of $F(\vec{y})$ only differ by an asymptotically vanishing first-order term. Furthermore, (\ref{mGPD}) may be combined with the empirical distribution function $\hat{F}(\vec{y})$ of $\vec{Y}_1,\ldots,\vec{Y}_n$ to provide an approximation to the full distribution of $F(\vec{y})$, namely
\begin{equation}\label{FullDistribution}
\hat{\hat{F}}(\vec{y})=\left\{\begin{array}{ll}
\hat{F}\{\min(\vec{y},\vec{u})\} + V[\min\{\tilde{t}_n(\vec{y}),\tilde{t}_n(\vec{u})\}] - V\{\tilde{t}_n(\vec{y})\}, & \vec{y}\nleq\vec{u},\\
\hat{F}(\vec{y}),&\vec{y}\leq \vec{u}.
\end{array}
\right.
\end{equation}
These approximations may be used with the probability integral transform to convert the data to the unit Fr\'echet scale as, e.g., in \citet{Coles.Tawn:1994}, \citet{Joe.etal:1992} and \citet{Huser.Davison:2014a}. Specifically, defining $\tilde{\tilde{t}}(\vec{y}):\Real^D\to\Real^D$ as the function such that $\tilde{\tilde{t}}(\vec{y})=\{\tilde{\tilde{t}}_1(y_1),\ldots,\tilde{\tilde{t}}_D(y_D)\}^T$ with
\begin{equation}\label{MarginalTransformation2}
\tilde{\tilde{t}}_d(y)=-1/\log\{\hat{\hat{F}}_d(y)\},\quad d=1,\ldots,D,
\end{equation}
and letting $\hat{\hat{F}}_d$ denote the $d$th marginal approximation in \eqref{FullDistribution}, one has that $\tilde{\tilde{t}}(\vec{y})\approx \tilde{t}_n(\vec{y})$ for $\vec{y}>\vec{u}$, and $\pr\{\tilde{\tilde{t}}_d(Y_d)\leq y\}\approx\exp(-1/y)$, $y>0$.

\subsection{The logistic model}
\label{ParametricModelsSection}
Although the marginal distributions in (\ref{GEV}) and (\ref{GPD}) depend on a finite number of parameters, the multivariate extreme-value and multivariate GPD distributions (\ref{ResultMaxima}) and (\ref{mGPD2}) are non-parametric because the underlying exponent measure $V(\vec{z})$ may be expressed in terms of a spectral measure taking almost any form; recall (\ref{ExponentMeasure}). In other words, there exists an infinite number of possible dependence structures for extremes. Classical inference relies on parametric families of exponent measures \citep[see, e.g.,][]{Tawn:1988b,Husler.Reiss:1989,Joe:1990,Coles.Tawn:1991,Joe.etal:1992,Boldi.Davison:2007,Ballani.Schlather:2011,Segers:2012,Sabourin.Naveau:2014}, and this section describes a well-established example, the logistic model, which we use in \S\ref{PerformanceAssessmentSection} to provide insight into the performance of different estimation procedures. 

The logistic model originates from \citet{Gumbel:1961} and puts
\begin{equation}\label{LogisticModel}
V(\vec{z}^\star)=\left(\sum_{d=1}^D {z_d^\star}^{-1/\alpha}\right)^{\alpha},\qquad \alpha\in(0,1].
\end{equation}
The limiting case $\alpha=1$ corresponds to independence, whereas the case $\alpha\to0$ corresponds to perfect dependence. In practice, this model suffers from a lack of flexibility, especially for large $D$, because the dependence structure is symmetric and summarized by a single parameter. A generalization that can capture non-exchangeability is the asymmetric logistic model proposed by \citet{Tawn:1988b} and \citet{Coles.Tawn:1991}, studied by \citet{Stephenson:2009}, and used by \citet{Ferrez.etal:2011} among others. The exponent measure may be expressed as
\begin{equation}\label{AsymmetricLogisticModel}
V(\vec{z}^\star)=\sum_{E\in\calE}\left\{\sum_{d\in E} \left({{z_d^\star}\over \theta_{E,d}}\right)^{-1/\alpha_E}\right\}^{\alpha_E},
\end{equation}
where $\calE$ is the set of all non-empty subsets of $\calD=\{1,\ldots,D\}$. The dependence parameters must satisfy $\alpha_E\in(0,1]$ for all sets $E\in\calE$ with $|E|>1$, and $\theta_{E,d}\in[0,1]$ with $\sum_{E\in\calE_{(d)}}\theta_{E,d}=1$ $(d=1,\ldots,D)$, where $\calE_{(d)}=\{E\in\calE:d\in E\}$. When $\alpha_{\calD}=\alpha$, $\theta_{\calD,d}=1$ and $\theta_{E,d}=0$ for all $d=1,\ldots,D$, $E\in\calE\setminus\calD$, the model (\ref{AsymmetricLogisticModel}) reduces to (\ref{LogisticModel}). As \citet{Stephenson:2009} pointed out, the full form of (\ref{AsymmetricLogisticModel}) is over-parametrized, but in practice simpler sub-models may be of interest. For example, \citet{Reich.Shaby:2012a} have shown that a model closely related to (though not a restriction of) (\ref{LogisticModel}) and (\ref{AsymmetricLogisticModel}) describes the finite-dimensional distributions of a particular max-stable spatial process, which they fit to precipitation extremes from a regional climate model. Furthermore, \citeauthor{Reich.Shaby:2012a}'s model converges in a certain sense to the \citet{Smith:1990b} model, which has been widely applied in the spatial extremes literature. Hence, although the model (\ref{LogisticModel}) is too rigid in most applications, it is closely related to more realistic settings, and, as such, is used as a model of reference in the present paper.

Generating data from models (\ref{LogisticModel}) and (\ref{AsymmetricLogisticModel}) can be easily and quickly performed in any dimension, thanks to their useful representations in terms of $\alpha$-stable variates \citep{Stephenson:2009}. 

In the following section, we present the main approaches to parametric inference based on the asymptotic results of \S\ref{AsymptoticTheorySection}.

\section{Inference}
\label{InferenceSection}
We now introduce several block maximum or threshold likelihood estimators that we shall compare in \S\ref{PerformanceAssessmentSection}. 
Suppose that the assumptions of result (\ref{ResultMaxima}) hold, and that $n=LN$ independent observations $\vec{y}_1,\ldots,\vec{y}_n$ distributed as the random vector $\vec{Y}$ have been recorded. The classical approach to inference is to form $N$ blocks of length $L$ with corresponding componentwise maxima $\vec{m}_1,\ldots,\vec{m}_N$ and to approximate the joint distribution of the latter by a parametric family of multivariate extreme-value distributions $G(\vec{z})=\exp[-V\{t_L(\vec{z});\psi\}]$, where $\psi\in\Psi\subset\Real^q$ denotes the vector of unknown marginal and dependence parameters. Here it is implicitly assumed that the transformation $t_L(\vec{z})$, defined in \eqref{MarginalTransformation} and \eqref{GEV}, involves location, scale and shape parameters to be estimated. This yields the log-likelihood function
\begin{equation}\label{LoglikelihoodMax1}
\ell_{{\rm Max},1}(\psi)=\sum_{i=1}^N\log\left(\sum_{P\in\calP}\prod_{E\in P}\left[-V_E\left\{t_L(\vec{m}_i);\psi\right\}\right]\right) - V\left\{t_L(\vec{m}_i);\psi\right\} + \log|J_{t_L}(\vec{m}_i)|,
\end{equation}
where $\calP$ is the collection of all partitions of $\calD=\{1,\ldots,D\}$, $V_E$ denotes the partial derivative of the function $V$ with respect to the variables whose indices lie in $E\subset\calD$, and $J_{t_L}(\vec{z})$ is the Jacobian associated with the transformation $t_L(\vec{z})$. Since the size of the set $\calP$ grows at a combinatorial rate as $D$ increases, \citet{Stephenson.Tawn:2005} proposed an alternative likelihood, which uses the extra information of occurrence times of maxima. More precisely, for each $i=1,\ldots,N$, let $P_{i}\subset\calP$ denote the partition that classifies block maxima $\vec{m}_i=(m_{i,1},\ldots,m_{i,D})^T$ according to their occurrence times, e.g., for $D=3$, if $m_{i,1}$ and $m_{i,2}$ occurred simultaneously, but separately from $m_{i,3}$, then $P_{i}=\{\{1,2\},\{3\}\}$. The Stephenson--Tawn log-likelihood may be written as
\begin{equation}\label{LoglikelihoodMax2}
\ell_{{\rm Max},2}(\psi)=\sum_{i=1}^N\sum_{E\in P_i}\log\left[-V_E\left\{t_L(\vec{m}_i);\psi\right\}\right] - V\left\{t_L(\vec{m}_i);\psi\right\} + \log|J_{t_L}(\vec{m}_i)|,
\end{equation}
thereby dramatically decreasing the number of terms in the log-likelihood. Recently, \citet{Wadsworth:2015} proposed a second-order bias correction of the Stephenson--Tawn likelihood, which may be written as
\begin{eqnarray}
\ell_{{\rm Max},3}(\psi)&=&\sum_{i=1}^N\log\bigg(\prod_{E\in P_i}\left[-V_E\left\{t_L(\vec{m}_i);\psi\right\}\right]\left\{1-{|P_i|(|P_i|-1)\over2L}\right\}\label{LoglikelihoodMax3}\\
&&+{1\over L}\sum_{\tilde{P}\prec P_i}\prod_{\tilde{E}\in \tilde{P}}\left[-V_{\tilde{E}}\left\{t_L(\vec{m}_i);\psi\right\}\right]\bigg) - V\left\{t_L(\vec{m}_i);\psi\right\} + \log|J_{t_L}(\vec{m}_i)|,\nonumber
\end{eqnarray}
where $|P_i|$ is the cardinality of the partition $P_i$, and $\tilde{P}\prec P_i$ denotes a sub-partition $\tilde{P}\in\calP$ of $P_i$ with cardinality $|\tilde{P}|=|P_i|-1$. This reduces the bias, while retaining a fairly small number of likelihood terms compared to \eqref{LoglikelihoodMax1}, at least in weak dependence scenarios. Another way to reduce the computational burden of (\ref{LoglikelihoodMax1}) is through composite likelihoods; see, e.g., \citet{Lindsay:1988}, \citet{Varin.Vidoni:2005}, or \citet{Varin.etal:2011}. In particular, pairwise likelihoods are constructed by multiplying all bivariate contributions, possibly weighted, under the working assumption of mutual independence. A log-pairwise likelihood based on block maxima may be written as
\begin{equation}\label{LoglikelihoodMaxPairwise}
\ell_{{\rm Max},{\rm Pair}}(\psi)=\sum_{i=1}^N\sum_{d_1<d_2} \log\left\{g\left(m_{i,d_1},m_{i,d_2};\psi\right)\right\},
\end{equation}
where $g(z_1,z_2;\psi)$ denotes the bivariate density stemming from $G(\vec{z})$ \citep{Padoan.etal:2010,Davison.Gholamrezaee:2012}. Maximum composite likelihood estimators and classical maximum likelihood estimators share similar asymptotic properties: both are strongly consistent, asymptotically Gaussian and converge at rate $\sqrt{N}$. However, the former are more variable than the latter, and require a special treatment of uncertainty \citep{Cox.Reid:2004,Padoan.etal:2010,Davis.Yau:2011,Huser.Davison:2013a}. 

More efficient inference can be performed using threshold methods. These primarily differ in the way threshold exceedances are defined and how they enter into the likelihood function. The first approach, developed by \citet{Coles.Tawn:1991}, consists in choosing a high \emph{marginal} threshold $\vec{u}\in\Real_+^D$ and building a likelihood from the Poisson process approximation (\ref{PointProcess}) for events falling in the extreme set $A_{\vec{u}}=[-\pmb{\infty},\pmb{\infty})\setminus[-\pmb{\infty},\vec{u}]$, i.e., whenever at least one variable exceeds its marginal threshold. If the threshold $\vec{u}$ is extreme enough, then exceedances over $\vec{u}$ should be approximately distributed according to a Poisson point process with intensity $\nu({\rm d}\vec{y})=-|J_{t_n}(\vec{y})|V_{1:D}\{t_n(\vec{y});\psi\}{\rm d}\vec{y}$, where $J_{t_n}(\vec{y})$ is the Jacobian associated with the transformation $t_n(\vec{y})$ defined in \eqref{MarginalTransformation} and \eqref{GEV}. Let $\vec{y}^i\in A_{\vec{u}}$, $i=1,\ldots,N_{\vec{u}}$, denote these exceedances. The corresponding Poisson log-likelihood is
\begin{equation}\label{LoglikelihoodThr1}
\ell_{{\rm Thr},1}(\psi)=-V\{t_n(\vec{u});\psi\}+\sum_{i=1}^{N_{\vec{u}}}\log\left[-V_{1:D}\left\{t_n(\vec{y}^i);\psi\right\}\right] + \log|J_{t_n}(\vec{y}^i)|.
\end{equation}
A second approach is to define extreme events as the observations $\vec{y}^i\in A^{\vec{r}}$, $i=1,\ldots,N^{\vec{r}}$, whose radial part exceeds a specific high \emph{diagonal} threshold vector $\vec{r}=(r_{1},\ldots,r_{D})^T$. Thanks to (\ref{ExponentMeasure2}), the corresponding Poisson log-likelihood is 
\begin{equation}\label{LoglikelihoodThr2}
\ell_{{\rm Thr},2}(\psi)\equiv\sum_{i=1}^{N^{\vec{r}}}\log\left[-V_{1:D}\left\{t_n(\vec{y}^i);\psi\right\}\right] + \log|J_{t_n}(\vec{y}^i)|,
\end{equation}
where $\equiv$ means equality up to an additive constant. A third approach is to use a likelihood constructed from the asymptotic multivariate GPD characterization; recall (\ref{mGPD}). Given a high marginal threshold $\vec{u}$ with corresponding exceedances $\vec{y}^i\in A_{\vec{u}}$, $i=1,\ldots,N_{\vec{u}}$, the log-likelihood function based on (\ref{mGPD2}) is
\begin{equation}\label{LoglikelihoodThr3}
\ell_{{\rm Thr},3}(\psi)=-N_{\vec{u}}\log\left[V\{t_n(\vec{u});\psi\}\right]+\sum_{i=1}^{N_{\vec{u}}}\log\left[-V_{1:D}\left\{t_n(\vec{y}^i);\psi\right\}\right] + \log|J_{t_n}(\vec{y}^i)|.
\end{equation}
For large $\vec{u}$, the variable $N_{\vec{u}}$ should be approximately distributed as a Poisson random variable with mean $V\{t_n(\vec{u});\psi\}$. If so, it turns out that $\ell_{{\rm Thr},1}(\psi)\equiv\ell_{{\rm Thr},3}(\psi)+\ell_{N_{\vec{u}}}(\psi)$ for any $\psi\in\Psi$, where $\ell_{N_{\vec{u}}}(\psi)$ is the log-likelihood for $N_{\vec{u}}$. This implies that the corresponding Fisher information matrices satisfy $I_{{\rm Thr},1}(\psi)=I_{{\rm Thr},3}(\psi)+I_{N_{\vec{u}}}>I_{{\rm Thr},3}(\psi)$, so that inference based on the log-likelihood (\ref{LoglikelihoodThr1}) is more efficient than using (\ref{LoglikelihoodThr3}). In fact, (\ref{LoglikelihoodThr1}) treats the number of exceedances as random, whereas (\ref{LoglikelihoodThr3}) conditions upon it. However, the improvement of (\ref{LoglikelihoodThr1}) over (\ref{LoglikelihoodThr3}) is slight as $n\to\infty$. \citet{Michel:2009} proposed alternative efficient likelihood procedures for multivariate GPD data.

The likelihoods \eqref{LoglikelihoodThr1}, \eqref{LoglikelihoodThr2} and \eqref{LoglikelihoodThr3} require that all the mass of the exponent measure is distributed on the interior of its domain of definition (as for the logistic model), and are unsuitable if some positive mass lies on the boundary faces or edges (as for the asymmetric logistic model); see \citet{Thibaud.Opitz:2015}.

\begin{figure}[t!]
  \includegraphics[width=4.55in]{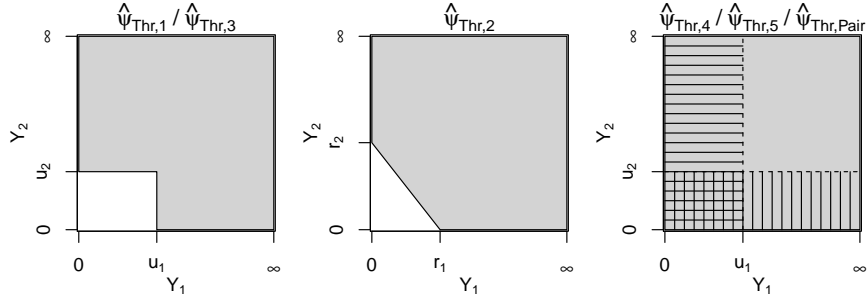}
\caption{Schematic view of the different threshold-based approaches to inference, illustrated for $D=2$ and with unit Fr\'echet margins. \emph{Left}: Poisson likelihood with marginal thresholds and multivariate GPD approach; \emph{middle}: Poisson likelihood with diagonal threshold; \emph{right}: censored likelihood approaches. Data points lying in the grey areas contribute to the likelihood, and censoring is indicated with shaded lines.}
\label{Fig:IllustrationThresholds} 
\end{figure}

A fourth approach, which works also for models with mass on boundary faces or edges, is to approximate the joint distribution $F(\vec{y})$ by using (\ref{ApproximationTail}) or (\ref{ApproximationTail2}) and to adopt a censored approach to account for misspecification below a high marginal threshold $\vec{u}$. To be more precise, let $\pmb{\delta}_i\in\{0,1\}^D$ $(i=1,\ldots,n)$ denote indicator variables reporting whether $y_{i,d}>u_d$ ($\delta_{i,d}=1$) or $y_{i,d}\leq u_d$ ($\delta_{i,d}=0$). Each observation $\vec{y}_i$ can then be split into a vector of exceedances, $\vec{y}_i^>$, and a vector of non-exceedances, $\vec{y}_i^\leq$. The censoring scheme that we consider supposes that the available set of observations is composed of $(\pmb{\delta}_i,\vec{y}_i^>)$ $(i=1,\ldots,n)$. Further, define the vectors $\vec{u}_i^>$ and $\vec{u}_i^\leq$, containing the elements of the threshold vector $\vec{u}$ corresponding to exceedances and non-exceedances. Then, if $F(\vec{y})$ is a suitable model for $\vec{y}>\vec{u}$, the contribution to the likelihood of a censored observation $(\pmb{\delta}_i,\vec{y}_i^>)$ is
\begin{equation}\label{Censored}
p_{\vec{u}}(\vec{y}_i;\psi)=\int_{-\pmb{\infty}}^{\vec{u}_i^\leq} {\rm d}F(\vec{y}_i) {\rm d}\vec{y}_i^\leq = F_{\pmb{\delta}_i}(\vec{b}_i),
\end{equation}
where the vector $\vec{b}_i$ has components $b_{i,d}=\max(y_{i,d},u_d)$, and where $F_{\pmb{\delta}_i}(\vec{y})$ denotes partial differentiation of the distribution $F(\vec{y})$ with respect to the variables corresponding to $\delta_{i,d}=1$ $(d=1,\ldots,D)$. Approximations $p^1_{\vec{u}}(\vec{y}_i;\psi)$ and $p^2_{\vec{u}}(\vec{y}_i;\psi)$ to (\ref{Censored}) may be obtained by replacing the distribution $F$ by the tail approximations in the right-most expression of (\ref{ApproximationTail}) and (\ref{ApproximationTail2}), respectively. Summing up all log-censored contributions, we get the log-likelihood functions
\begin{equation}\label{LoglikelihoodThr4:5}
\ell_{{\rm Thr},4}(\psi)=\sum_{i=1}^n \log\left\{p^1_{\vec{u}}(\vec{y}_i;\psi)\right\},\quad \ell_{{\rm Thr},5}(\psi)=\sum_{i=1}^n \log\left\{p^2_{\vec{u}}(\vec{y}_i;\psi)\right\}.
\end{equation}
The censored likelihood $\ell_{{\rm Thr},4}(\psi)$ was proposed by \citet{Ledford.Tawn:1996} and applied in the bivariate case by \citet{Bortot.etal:2000} and \citet[][p.155]{Coles:2001}, while $\ell_{{\rm Thr},5}(\psi)$ was advocated by \citet{Smith.etal:1997} and recently extended to the spatial framework by \citet{Wadsworth.Tawn:2014} and \citet{Thibaud.Opitz:2015}, albeit with a slight modification for the points falling in $[-\pmb{\infty},\vec{u}]$. When the exponent measure or its partial derivatives are not available for $D>2$, and to reduce the computational burden, \citet{Thibaud.etal:2013} and \citet{Huser.Davison:2014a} propose a censored pairwise likelihood similar to
\begin{equation}\label{LoglikelihoodThrPairwise}
\ell_{{\rm Thr},{\rm Pair}}(\psi)=\sum_{i=1}^n\sum_{d_1<d_2} \log\left\{p^1_{\vec{u}}(y_{i,d_1},y_{i,d_2};\psi)\right\},
\end{equation}
where $p^1_{\vec{u}}(y_1,y_2;\psi)$ is the bivariate counterpart of $p^1_{\vec{u}}(\vec{y};\psi)$; see also \citet{Bacro.Gaetan:2014}. Alternatively, a partially censored pairwise likelihood was proposed by \citet{Wadsworth.Tawn:2012b}. The domains of these different threshold-based estimators are illustrated in Figure~\ref{Fig:IllustrationThresholds}.

The notation $\hat\psi_{\cdot,j}=\arg\max_{\psi\in\Psi}\ell_{\cdot,j}$, $\hat\psi_{\cdot,{\rm Pair}}=\arg\max_{\psi\in\Psi}\ell_{\cdot,{\rm Pair}}$ is used hereafter to denote maximum likelihood estimators and maximum pairwise likelihood estimators, respectively. In \S\ref{PerformanceAssessmentSection}, we compute their asymptotic relative efficiencies for the logistic model when $D=2$, and assess their empirical performance for $D\geq2$. 

\section{Performance assessment of estimators}
\label{PerformanceAssessmentSection}

\subsection{Two-dimensional case}
\label{TwoDimensionalCaseSection}

\paragraph{Asymptotic relative efficiencies.}
\label{AsymptoticRelativeEfficienciesSection}

Since under standard regularity conditions, maximum (composite) likelihood estimators are asymptotically unbiased (see, e.g., \citealp{Davison:2003}, p.122--125, and \citealp{Varin.etal:2011}), their asymptotic relative efficiency, i.e., the ratio of variances as $n\to\infty$, is a natural measure of performance. In dimension $D=2$, maximum pairwise likelihood estimators coincide with their full likelihood counterparts, whose variance equals the reciprocal Fisher information, as $n\to\infty$. The latter was worked out for the multivariate logistic model with unknown marginals by \citet{Shi:1995a}. Furthermore, \citet{Stephenson.Tawn:2005} investigated the asymptotic relative efficiency of $\hat\psi_{{\rm Max},2}$ with respect to $\hat\psi_{{\rm Max},1}$ based on (\ref{LoglikelihoodMax2}) and (\ref{LoglikelihoodMax1}) for the bivariate logistic model with known margins; the same calculations apply to the second-order bias reduction approach $\hat\psi_{{\rm Max},3}$. However, nobody has yet assessed the asymptotic variance of threshold estimators based on the Poisson likelihood or the censored likelihood. Here, we compute it theoretically for $\hat\psi_{{\rm Thr},4}$ and $\hat\psi_{{\rm Thr},5}$ (see the Appendix for the details) and by simulation for $\hat\psi_{{\rm Thr},1}$, $\hat\psi_{{\rm Thr},2}$ and $\hat\psi_{{\rm Thr},3}$ in the case of the bivariate logistic model with known unit Fr\'echet margins, i.e., with marginal transformations (\ref{GEV}) and (\ref{MarginalTransformation2}) satisfying $t_n(\vec{z})=\tilde{t}_n(\vec{z})=\tilde{\tilde{t}}(\vec{z})=\vec{z}$. Accordingly, the notation $\hat\alpha$, with subscripts consistent with \S\ref{InferenceSection}, will be used instead of $\hat\psi$. Block maximum estimators assume a block length $L=100$, while threshold estimators are defined in terms of the threshold probability $p$: the marginal threshold $\vec{u}(p)$ is chosen as the vector of $p$-quantiles, while the diagonal threshold $\vec{r}(p)=\{r(p),\ldots,r(p)\}^T$ is such that there are $100\times(1-p)\%$ exceedances over it on average. Table~\ref{AsymptoticRelativeEfficienciesTable} reports the (theoretical or simulation-based) root asymptotic relative efficiencies of the estimators with respect to the censored likelihood estimator $\hat\alpha_{{\rm Thr},4}$. Results obtained by simulation are based on $10^5$ independent datasets simulated from the logistic model with sample size $n=50000$.

\begin{table}[t!]
\caption{Root asymptotic relative efficiencies (\%) of the different estimators of $\alpha$ introduced in \S\ref{InferenceSection} with respect to the censored estimator $\hat\alpha_{{\rm Thr},4}$ with threshold probability $p=0.95$ (first four rows) and $p=0.99$ (last four rows), for the bivariate logistic model (\ref{LogisticModel}) and different values of the dependence parameter $\alpha$. In this bivariate setting, $\hat\alpha_{{\rm Max},{\rm Pair}}= \hat\alpha_{{\rm Max},1}$ and $\hat\alpha_{{\rm Thr},{\rm Pair}}= \hat\alpha_{{\rm Thr},4}$. Moreover, $\hat\alpha_{{\rm Max},3}$ is asymptotically as efficient as $\hat\alpha_{{\rm Max},2}$, and similarly for $\hat\alpha_{{\rm Thr},{\rm 3}}$ with respect to $\hat\alpha_{{\rm Thr},{\rm 1}}$, and $\hat\alpha_{{\rm Thr},{\rm 5}}$ with respect to $\hat\alpha_{{\rm Thr},{\rm 4}}$. For block maximum estimators, the number of observations per block is set to $L=100$.}
\label{AsymptoticRelativeEfficienciesTable}
\begin{tabular}{llrrrrrrrrr}
\hline\noalign{\smallskip}
Estim. & $L$ or $p$ & \multicolumn{9}{c}{Dependence parameter $\alpha$}  \\
& & $0.1$ & $0.2$ & $0.3$ & $0.4$ & $0.5$ & $0.6$ & $0.7$ & $0.8$ & $0.9$\\
\noalign{\smallskip}\hline\noalign{\smallskip}
$\hat\alpha_{{\rm Max},1}{}^{\dagger}$  & $L=100$ & $42.6$ & $40.4$ & $38.1$ & $35.8$ & $33.3$ & $30.7$ & $28.0$ & $24.8$ & $20.8$ \\
$\hat\alpha_{{\rm Max},2}{}^{\dagger}$ & $L=100$ & $44.2$ & $43.8$ & $43.5$ & $43.3$ & $43.1$ & $43.0$ & $43.1$ & $43.4$ & $45.3$ \\
$\hat\alpha_{{\rm Thr},1}{}^{\ddagger}$ & $p=0.95$ & $108.0$ & $117.0$ & $128.0$ & $145.0$ & $167.0$ & $199.0$ & $237.0$ & $275.0$ & $291.0$ \\
$\hat\alpha_{{\rm Thr},2}{}^{\ddagger}$ & $p=0.95$ & $99.0$ & $100.0$ & $103.0$ & $111.0$ & $124.0$ & $146.0$ & $177.0$ & $215.0$ & $248.0$ \\
\noalign{\smallskip}\hline\noalign{\smallskip}
$\hat\alpha_{{\rm Max},1}{}^{\dagger}$ & $L=100$ & $95.2$ & $90.1$ & $84.9$ & $79.4$ & $73.6$ & $67.5$ & $60.7$ & $52.7$ & $41.9$ \\
$\hat\alpha_{{\rm Max},2}{}^{\dagger}$ & $L=100$ & $98.8$ & $97.8$ & $96.9$ & $96.1$ & $95.3$ & $94.5$ & $93.5$ & $92.2$ & $91.3$ \\
$\hat\alpha_{{\rm Thr},1}{}^{\ddagger}$ & $p=0.99$ & $107.0$ & $116.0$ & $126.0$ & $139.0$ & $158.0$ & $188.0$ & $232.0$ & $293.0$ & $348.0$ \\
$\hat\alpha_{{\rm Thr},2}{}^{\ddagger}$ & $p=0.99$ & $99.0$ & $100.0$ & $102.0$ & $108.0$ & $118.0$ & $137.0$ & $168.0$ & $218.0$ & $278.0$ \\
\noalign{\smallskip}\hline
\multicolumn{11}{l}{${}^{\dagger}$: {\scriptsize Numbers calculated theoretically.}}  \\
\multicolumn{11}{l}{${}^{\ddagger}$: {\scriptsize Numbers calculated by simulation from $10^5$ estimates of $\alpha$ obtained from bivariate logistic data of size $50000$.}}  \\
\end{tabular}
\end{table}

As expected, threshold-based estimators outperform block maximum estimators. But, more interestingly, the former are less variable than the latter even when the same number of ``useful'' observations is available for both estimation procedures ($L=100$ and $p=0.99$). Surprisingly, this discrepancy increases as $\alpha$ approaches unity, where data are closer to independence, and so are more likely to be censored using $\hat\alpha_{{\rm Thr},{\rm 4}}$. By contrast, the effect of censoring in $\hat\alpha_{{\rm Thr},{\rm 4}}$ is striking when considering the relative efficiency with respect to non-censored threshold estimators $\hat\alpha_{{\rm Thr},{\rm 1}}$ and $\hat\alpha_{{\rm Thr},{\rm 2}}$. The latter, which use the actual values of additional data points close to the axes (recall Figure~\ref{Fig:IllustrationThresholds}), increasingly outperform the censored likelihood estimator as $\alpha\to1$. For example, when $\alpha=0.9$, the asymptotic standard deviation of the censored estimator $\hat\alpha_{{\rm Thr},{\rm 4}}$ is almost three and a half times that of $\hat\alpha_{{\rm Thr},{\rm 1}}$ at the $99\%$ threshold. This suggests that censoring discards non-negligible information when the data are nearly independent. However, at sub-asymptotic regimes with finite $n$, the biases and robustness of these estimators should also be taken into account. In particular, if block sizes (respectively thresholds) are not large enough, the approximation of block maxima (respectively threshold exceedances) by their asymptotic distribution might induce some misspecification bias. We assess this by simulation in dimension $D=2$.

\paragraph{Estimation ability.}
\label{EstimationAbilitySection}

\begin{figure}
  \includegraphics[width=\linewidth]{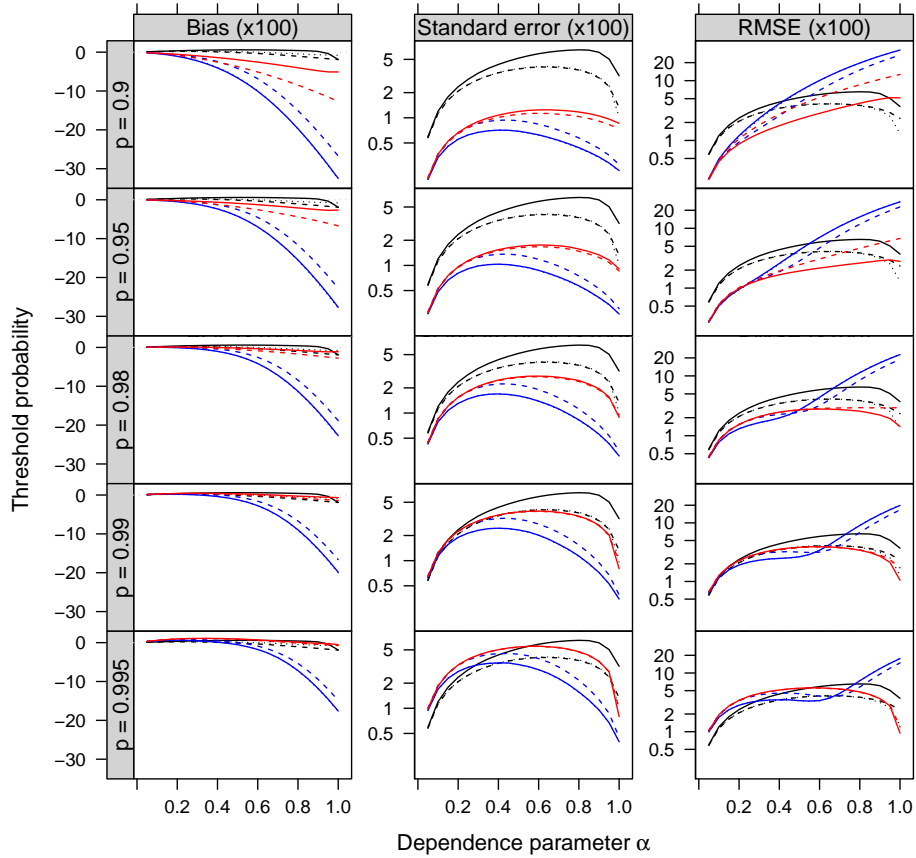}
\caption{Empirical bias (left), standard error (middle) and root mean squared error (right) of the different estimators introduced in \S\ref{InferenceSection}, to assess the dependence strength of the limiting bivariate logistic model, plotted against the true dependence parameter $\alpha$ and for threshold probabilities $p=0.9,0.95,0.98,0.99,0.995$ (top to bottom rows). Block-maximum estimators correspond to black curves ($\hat\alpha_{{\rm Max},{\rm 1}}$ solid, $\hat\alpha_{{\rm Max},{\rm 2}}$ dashed, $\hat\alpha_{{\rm Max},{\rm 3}}$ dotted), while Poisson likelihood or multivariate GPD-based estimators are in blue ($\hat\alpha_{{\rm Thr},{\rm 1}}$ solid, $\hat\alpha_{{\rm Thr},{\rm 2}}$ dashed, $\hat\alpha_{{\rm Thr},{\rm 3}}$ dotted), and censored estimators are in red ($\hat\alpha_{{\rm Thr},{\rm 4}}$ solid, $\hat\alpha_{{\rm Thr},{\rm 5}}$ dashed). Estimators $\hat\alpha_{{\rm Thr},{\rm 1}}$ and $\hat\alpha_{{\rm Thr},{\rm 3}}$ are almost indistinguishable, and similarly for $\hat\alpha_{{\rm Max},{\rm 2}}$ and $\hat\alpha_{{\rm Max},{\rm 3}}$. In this bivariate setting, $\hat\alpha_{{\rm Max},{\rm Pair}}= \hat\alpha_{{\rm Max},1}$ and $\hat\alpha_{{\rm Thr},{\rm Pair}}= \hat\alpha_{{\rm Thr},4}$. $R=10^4$ independent replicates were used to compute these values. Standard errors and RMSE are displayed on a logarithmic scale.}
\label{Fig:SimulationStudy1} 
\end{figure}

In order to assess the practical performance of the different methods introduced in \S\ref{InferenceSection} in terms of bias and efficiency, we conducted a simulation study, in which data were generated in the max-domain of attraction of the logistic model (\ref{LogisticModel}). For different values of $\alpha$ ranging from very strong dependence ($\alpha=0.05$) to independence ($\alpha=1$), we simulated $R=10^4$ independent datasets of size $n=10^4$ from an Archimedean copula with generator $\varphi(t)=(t^\alpha+1)^{-1}$ (known as the outer power Clayton copula, see \citealp{Hofert.etal:2015} and \citealp{Nelsen:2006}) and zero-truncated Student $t$ marginals. In other words, the joint distribution function $F(\vec{y})$ of our simulated observations is 
\begin{equation}\label{Copula}
F(\vec{y})=\varphi\left[\varphi^{-1}\left\{F_1(y_1)\right\}+\cdots+\varphi^{-1}\left\{F_D(y_D)\right\}\right],
\end{equation}
where for each $d=1,\ldots,D$, the marginal distributions satisfy $F_d(0)=0.5$ and  $F_d(y)=0.5 + 0.5T_5(y)$, $y>0$, with $T_5(y)$ denoting the $t$ distribution function with $5$ degrees of freedom. The simulated data are positive, with a positive mass at zero, and heavy-tailed, which are common features of rainfall data for example \citep{Huser.Davison:2014a}. The presence of the point mass at zero could be problematic for non-censored estimators. The distribution (\ref{Copula}) is known to be in the max-domain of attraction of the logistic model with GEV margins (\ref{GEV}) with shape parameters $\xi_d=0.2$ $(d=1,\ldots,D)$; see \citet{Fougeres:2004} and \citet[][p.59]{Beirlant.etal:2004}. For this simulation study, we focus on the bivariate case with $D=2$. In order to estimate the dependence parameter $\alpha$, we consider a two-step approach: First, once block maxima (respectively threshold exceedances) are identified, the GEV distribution (respectively the GPD (\ref{GPD})) is fitted to each margin separately. Second, the limiting logistic model is fitted using the different estimators of \S\ref{InferenceSection}, treating the estimated marginals as fixed. One-step estimators are also considered in the Supplementary Material. To quantify estimation ability, the $R$ replicates of each estimator considered are then used to compute its empirical bias, standard error and root mean squared error (RMSE). More precisely, denoting independent replicates of some estimator $\hat\alpha$ for $\alpha$ by $\hat\alpha_r$ $(r=1,\ldots,R)$, we define 
\begin{eqnarray}\label{BiasSeRMSE}
{\rm Bias}(\hat\alpha)&=&\bar{\hat\alpha}-\alpha,\quad{\rm SE}(\hat\alpha)\;\;=\;\;\left\{{1\over R-1}\sum_{r=1}^R\left(\hat\alpha_r-\bar{\hat\alpha}\right)^2\right\}^{1/2},\nonumber\\
{\rm RMSE}(\hat\alpha)&=&\left[\{{\rm Bias}(\hat\alpha)\}^2 + \{{\rm SE}(\hat\alpha)\}^2\right]^{1/2},
\end{eqnarray}
where $\bar{\hat\alpha}=R^{-1}\sum_{r=1}^R\hat\alpha_r$. As above, the block maximum estimators use block size $L=100$, so that $N=100$ maxima are available for fitting. In practice, this setting could correspond to $100$ summer maxima of data recorded on a daily basis. For threshold estimators, we consider threshold probabilities $p=0.9,0.95,0.98,0.99,0.995$. The results are reported in Figure~\ref{Fig:SimulationStudy1}.

Overall, the relative efficiencies are consistent with their asymptotic counterparts in Table~\ref{AsymptoticRelativeEfficienciesTable}, though with some slight differences due to the estimation of margins. However, this simulation study offers new insight for finite $n$: all estimators tend to overestimate the strength of dependence, and this overestimation increases as the data become more independent, i.e., as $\alpha$ approaches unity. As expected, block-maximum estimators have a limited bias and huge variability, though $\hat\alpha_{{\rm Max},{\rm 2}}$ and $\hat\alpha_{{\rm Max},{\rm 3}}$ outperform $\hat\alpha_{{\rm Max},{\rm 1}}$. In this bivariate setting, the bias-reduction estimator $\hat\alpha_{{\rm Max},{\rm 3}}$ behaves very similarly to its counterpart $\hat\alpha_{{\rm Max},{\rm 2}}$ but offers a slightly better performance close to independence. The former is comparable to the censored estimator $\hat\alpha_{{\rm Thr},{\rm 4}}$ at the $99\%$ level, where the number of exceedances is the same as the number of block maxima (using a block size $L=100$). Regarding threshold estimators with $p=0.9,0.95,0.98$, the best performance overall according to the RMSE is attained by the censored estimator $\hat\alpha_{{\rm Thr},{\rm 4}}$, whose increased variability compared to $\hat\alpha_{{\rm Thr},{\rm 1}}$, $\hat\alpha_{{\rm Thr},{\rm 2}}$, $\hat\alpha_{{\rm Thr},{\rm 3}}$ is compensated by a well-controlled bias. For higher thresholds, with $p=0.99,0.995$, estimators based on Poisson likelihoods perform slightly better when $\alpha<0.7$, which was expected since the limiting model is likely to fit better. Non-censored threshold-based estimators are fairly reliable for very high $p$ and small $\alpha$, a situation rarely encountered in practice, but perform very badly at moderate thresholds or when the data are nearly independent. They suffer from a pronounced bias owing to their sensitivity to model misspecification close to the axes, whereas censored or block-maximum estimators are more robust. Interestingly, although block maximum estimators with $L=100$ use about five times less data than non-censored estimators with $p=0.95$, the former nevertheless have lower RMSEs than the latter when $\alpha>0.5$. 

To summarize, at extreme levels often considered in practice and for a large range of dependence strengths, censored estimators, and especially $\hat\alpha_{{\rm Thr},{\rm 4}}$, seem to offer the best compromise between robustness (small bias) and efficiency (low variability). Table~\ref{BestEstimator1Table} summarizes the results of an extended simulation study, showing that the estimator $\hat\alpha_{{\rm Thr},4}$ is always found to be best, when the comparison is done across a wide range of threshold probabilities $p$ and block lengths $L$. Interestingly, as dependence decreases, the threshold considered should increase. This provides strong support for the use of the censored estimator $\hat\alpha_{{\rm Thr},4}$ in practice, and can guide the choice of the threshold probability. Similar results (not shown) were found for sample sizes $n=2000$ and $n=50000$. In the Supplementary Material, we also show that when marginal and dependence parameters are estimated simultaneously, similar conclusions hold, though the diagonal threshold estimator $\hat\alpha_{{\rm Thr},2}$ has an overall decreased performance.

We now investigate the predictive ability of these estimators in a similar setting.

\begin{table}[t!]
\caption{Best estimator overall in terms of RMSE for different dependence strengths. The results are based on a simulation study with sample size $n=10^4$ and dimension $D=2$, and the comparison is performed across several block maximum estimators with block length $L=20,50,100,200,500,1000$, and threshold-based estimators with threshold probability $p=0.9,0.95,0.96,0.97,0.98,0.99,0.995,0.999$. For more details, see \S\ref{InferenceSection} and \S\ref{AsymptoticRelativeEfficienciesSection}.}
\label{BestEstimator1Table}
\begin{tabular}{lllllllllll}
\hline\noalign{\smallskip}
 & \multicolumn{9}{c}{Dependence parameter $\alpha$}  \\
& $0.1$ & $0.2$ & $0.3$ & $0.4$ & $0.5$ & $0.6$ & $0.7$ & $0.8$ & $0.9$ & $1$\\
\noalign{\smallskip}\hline\noalign{\smallskip}
Estim. & $\hat\alpha_{{\rm Thr},4}$ & $\hat\alpha_{{\rm Thr},4}$ & $\hat\alpha_{{\rm Thr},4}$ & $\hat\alpha_{{\rm Thr},4}$ & $\hat\alpha_{{\rm Thr},4}$ & $\hat\alpha_{{\rm Thr},4}$ & $\hat\alpha_{{\rm Thr},4}$ & $\hat\alpha_{{\rm Thr},4}$ & $\hat\alpha_{{\rm Thr},4}$ & $\hat\alpha_{{\rm Thr},4}$ \\
$p$ & $0.9$ & $0.9$ & $0.9$ & $0.95$ & $0.95$ & $0.95$ & $0.96$ & $0.97$ & $0.98$ & $0.999$ \\
\noalign{\smallskip}\hline
\end{tabular}
\end{table}

\begin{figure}
  \includegraphics[width=\linewidth]{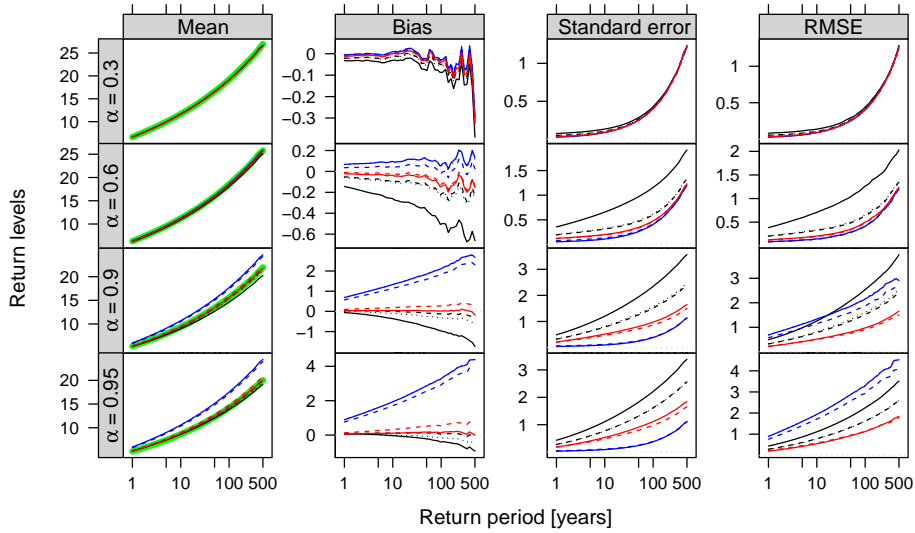}
\caption{Empirical mean, bias, standard error and RMSE (left to right columns) for return levels of the sum $Y_1+Y_2$ of a vector $\vec{Y}$ in the max-domain of attraction of the bivariate logistic model, displayed against the corresponding return periods (on a logarithmic scale), for strong dependence (top row), mild dependence (second row), weak dependence (third row) and near independence (bottom row) scenarios. The thick green curves represent the true return levels, while the other curves are obtained from $R=10^4$ independent estimates using various approaches. Block-maximum estimators correspond to black curves ($\hat\alpha_{{\rm Max},{\rm 1}}$ solid, $\hat\alpha_{{\rm Max},{\rm 2}}$ dashed, $\hat\alpha_{{\rm Max},{\rm 3}}$ dotted), while Poisson likelihood or multivariate GPD-based estimators are in blue ($\hat\alpha_{{\rm Thr},{\rm 1}}$ solid, $\hat\alpha_{{\rm Thr},{\rm 2}}$ dashed, $\hat\alpha_{{\rm Thr},{\rm 3}}$ dotted), and censored estimators are in red ($\hat\alpha_{{\rm Thr},{\rm 4}}$ solid, $\hat\alpha_{{\rm Thr},{\rm 5}}$ dashed).}
\label{Fig:SimulationStudy2} 
\end{figure}

\paragraph{Prediction.}
\label{PredictionAbilitySection}
In applications of extreme-value statistics, it is common to attempt to predict the largest event that might occur in a long future period, based on limited data. In order to assess how the estimators of \S\ref{InferenceSection} can predict the probabilities of such future extreme events, we conducted an additional simulation study in dimension $D=2$, based on the logistic model. In order to mimic a realistic setting, we simulated independent datasets from model (\ref{Copula}) with $n=20\times 100=2000$, which could be thought of as daily rainfall observations recorded during $20$ summers. For strong ($\alpha=0.3$), mild ($\alpha=0.6$), weak ($\alpha=0.9$) and very weak ($\alpha=0.95$) dependence, we estimated the dependence parameter $\alpha$ using the estimators previously described, and derived by simulation the return levels for the \emph{risk variable} $Y_1+Y_2$ based on the fitted value of $\alpha$ and the true marginals. \citet{Zheng.etal:2014} investigated alternative risk functions. We consider return periods ranging from $1$ up to $500$ years, which corresponds to an exceedance probability of $2\times10^{-5}$, i.e., once every $50000$ observations on average. We use annual block-maximum estimators (i.e., $L=100$) and set $p=0.98$ for threshold estimators, so that the latter use approximately twice as much data as the former. Repeating this procedure $R=10^4$ times, we then compile the independent replicates to compute the empirical mean, bias, standard error, and RMSE of the return levels; recall (\ref{BiasSeRMSE}). The results are reported in Figure~\ref{Fig:SimulationStudy2}.

Although these results do not reflect the real bias and uncertainty of return level estimators (because the latter were computed using the true marginals), we can use them to compare the performance of the different estimators in various dependence cases. For all estimators, the standard error increases drastically with the return period, as expected. The absolute bias also seems to increase, albeit at a slower rate. For strong to mild dependence scenarios with $\alpha=0.3,0.6$, all estimators perform quite well overall, though block maximum estimators are more variable than threshold estimators, and some slight positive (respectively negative) bias is observed for Poisson likelihood (respectively block-maximum) methods. In terms of RMSE, threshold-based estimators perform similarly, though Poisson likelihood methods are slightly better than censored methods, and they all outperform block-maximum estimators, especially $\hat\alpha_{{\rm Max},1}$. From weak dependence to near independence cases with $\alpha=0.9,0.95$, Poisson likelihood estimators are strongly positively biased, hence not reliable, block maximum estimators are very variable and slightly negatively biased, and censored estimators have good properties overall. When $\alpha=0.95$, block maximum estimators outperform Poisson likelihood estimators in terms of RMSE, though the latter use twice as much data as the former, and this improvement is likely to be more pronounced as $\alpha\to1$. Overall, the predictive ability of censored methods is much better than their competitors, especially in low dependence cases, and this improvement should be even more marked at lower thresholds.

\subsection{Performance in higher dimensions}
\label{PerformanceHigherDimensionsSection}
\begin{figure}[t!]
  \includegraphics[width=\linewidth]{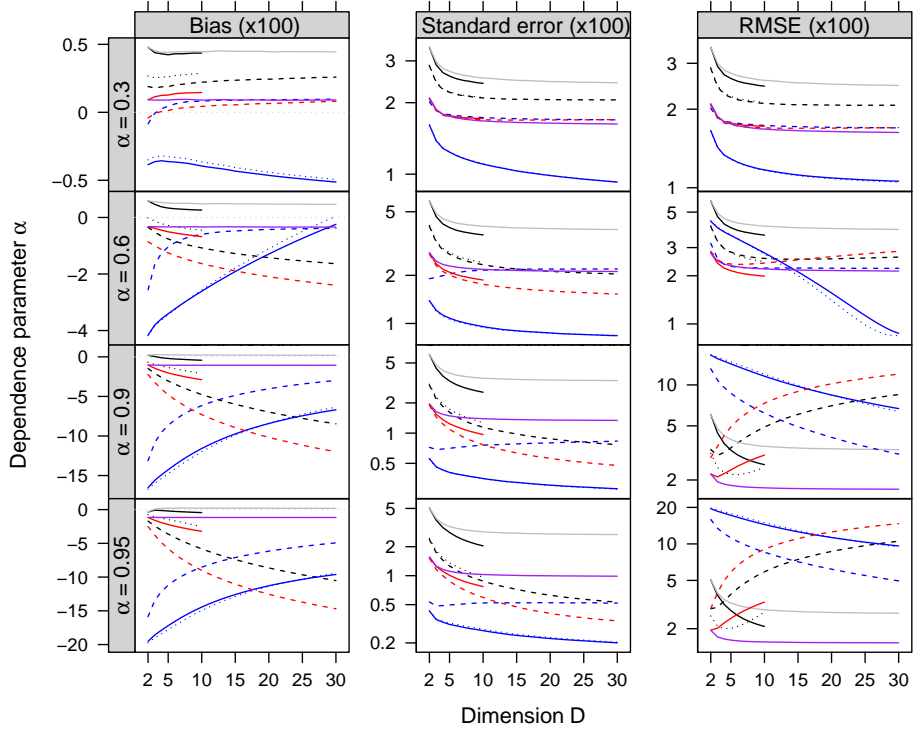}
\caption{Empirical bias (left), standard error (middle) and root mean squared error (right) of the different estimators of the dependence parameter $\alpha=0.3,0.6,0.9,0.95$ of the limiting logistic model, plotted against the dimension $D$. Block-maximum estimators correspond to black/grey curves ($\hat\alpha_{{\rm Max},{\rm 1}}$ black-solid, $\hat\alpha_{{\rm Max},{\rm 2}}$ black-dashed, $\hat\alpha_{{\rm Max},{\rm 3}}$ black-dotted, $\hat\alpha_{{\rm Max},{\rm Pair}}$ grey-solid), while the results for the Poisson likelihood or multivariate GPD-based estimators are in blue ($\hat\alpha_{{\rm Thr},{\rm 1}}$ solid, $\hat\alpha_{{\rm Thr},{\rm 2}}$ dashed, $\hat\alpha_{{\rm Thr},{\rm 3}}$ dotted), and those for censored estimators are in red/purple ($\hat\alpha_{{\rm Thr},{\rm 4}}$ red-solid, $\hat\alpha_{{\rm Thr},{\rm 5}}$ red-dashed, $\hat\alpha_{{\rm Thr},{\rm Pair}}$ purple-solid). Block lengths were set to $L=100$ and threshold probabilities to $p=0.98$. $R=10^4$ independent replicates were used to compute these values. Standard errors and RMSEs are displayed on a logarithmic scale.}
\label{Fig:SimulationStudy3} 
\end{figure}
In \S\ref{TwoDimensionalCaseSection}, we explored the performance of the different estimators of \S\ref{InferenceSection} for the bivariate logistic model. In order to understand how estimators compare in higher dimensions, we conducted an additional simulation study. In order to be consistent with \S\ref{TwoDimensionalCaseSection}, we generated independent $30$-dimensional datasets of size $n=10^4$ from the model (\ref{Copula}). We consider the cases of strong dependence ($\alpha=0.3$), mild dependence ($\alpha=0.6$), weak dependence ($\alpha=0.9$) and near independence ($\alpha=0.95$), and for each scenario we estimate the dependence parameter $\alpha$ with the different two-step estimators based on the $D=2,\ldots,30$ first components from the simulated data. As mentioned in \S\ref{InferenceSection}, the exact computation of $\hat\alpha_{{\rm Max},1}$  and $\hat\alpha_{{\rm Thr},4}$ is very demanding in high dimensions; the same is true for $\hat\alpha_{{\rm Max},3}$ when the dependence strength is strong. Monte Carlo approximations to the corresponding likelihood functions may be obtained for the logistic model based on the generation of a large number of $\alpha$-stable random variates \citep{Stephenson:2009,Fougeres.etal:2009,Huser:2013}, but this approach is difficult to apply in practice. Hence, for simplicity, we restrict ourselves to $D=2,\ldots,10$ in these cases. As above, block lengths are set to $L=100$ and threshold probabilities to $p=0.98$. 
We repeated this procedure $R=10^4$ times, in order to compute the empirical bias, standard error and RMSE for the different estimators considered; see (\ref{BiasSeRMSE}). The results are reported in Figure~\ref{Fig:SimulationStudy3}.

For the estimators $\hat\alpha_{{\rm Max},2}$, $\hat\alpha_{{\rm Max},3}$, $\hat\alpha_{{\rm Thr},4}$ and $\hat\alpha_{{\rm Thr},5}$ (i.e., Stephenson--Tawn-based and censored estimators), the absolute bias tends to increase with dimension, while the standard errors decrease. By contrast, for $\hat\alpha_{{\rm Thr},1}$, $\hat\alpha_{{\rm Thr},2}$ and $\hat\alpha_{{\rm Thr},3}$ (i.e., point process or multivariate GPD-based estimators), the absolute bias decreases sharply as a function of $D$ when $\alpha=0.6,0.9,0.95$. Regarding pairwise likelihood estimators, their bias is more or less constant and their standard error decreases more slowly than for their full-likelihood counterparts. In terms of RMSE, the best estimator overall appears to be the censored pairwise likelihood estimator $\hat\alpha_{{\rm Thr},{\rm Pair}}$ for weak dependence ($\alpha=0.9,0.95$) or mild dependence in moderate dimensions ($\alpha=0.6$, $D<15$) and the point process estimator with marginal threshold $\hat\alpha_{{\rm Thr},1}$ for strong dependence ($\alpha=0.3$) or mild dependence in large dimensions ($\alpha=0.6$, $D\geq 15$). While the performance of the censored pairwise likelihood estimator appears reasonable for any dependence strength and dimension, point process estimators have a very poor performance in low dependence cases for any dimension. To counteract the very strong bias of the latter, one should consider a higher threshold or use pairwise likelihood estimators, which are robust against misspecification of high-order interactions. Interestingly, block maximum estimators (especially $\hat\alpha_{{\rm Max},3}$) also seem to perform rather well when dependence is weak, but do poorly when $\alpha<0.9$.

A by-product of this simulation study is the relative efficiencies of pairwise likelihood estimators in an extreme-value context. This has already been investigated in different frameworks by \citet{Cox.Reid:2004}, \citet{Renard.etal:2004}, \citet{Hjort.Varin:2008} and \citet{Davis.Yau:2011}, among others. Table \ref{AsymptoticRelativeEfficienciesPairwiseTable} summarizes the results from the above simulation setting.
\begin{table}[t!]
\caption{Root asymptotic relative efficiencies (\%) of the two-step pairwise likelihood estimator $\hat\alpha_{{\rm Max},{\rm Pair}}$ (respectively $\hat\alpha_{{\rm Thr},{\rm Pair}}$) of $\alpha$ introduced in \S\ref{InferenceSection} with respect to $\hat\alpha_{{\rm Max},1}$ (respectively $\hat\alpha_{{\rm Thr},4}$), for the limiting $D$-dimensional logistic model with $D=2,\ldots,10$ and different values of the dependence parameter $\alpha$. The data were simulated according to model (\ref{Copula}), and the efficiencies computed based on $R=10^4$ replicates.}
\label{AsymptoticRelativeEfficienciesPairwiseTable}
\begin{tabular}{llrrrrrrrrr}
\hline\noalign{\smallskip}
 & \multicolumn{9}{c}{Dimension $D$}  \\
Estim. & $\alpha$ & $2$ & $3$ & $4$ & $5$ & $6$ & $7$ & $8$ & $9$ & $10$ \\
\noalign{\smallskip}\hline\noalign{\smallskip}
$\hat\alpha_{{\rm Max},{\rm Pair}}$ & $0.3$ & $100.0$ & $97.2$ & $96.0$ & $95.4$ & $95.2$ & $95.0$ & $95.1$ & $95.0$ & $94.9$ \\
& $0.6$ & $100.0$ & $95.4$ & $92.7$ & $91.2$ & $90.3$ & $89.5$ & $89.0$ & $88.6$ & $88.3$ \\
& $0.9$ & $100.0$ & $93.0$ & $87.9$ & $83.4$ & $80.3$ & $77.7$ & $75.6$ & $74.0$ & $72.4$ \\
& $0.95$ & $100.0$ & $96.0$ & $91.3$ & $86.7$ & $82.1$ & $78.7$ & $75.8$ & $73.5$ & $71.4$ \\
\noalign{\smallskip}\hline\noalign{\smallskip}
$\hat\alpha_{{\rm Thr},{\rm Pair}}$  & $0.3$ & $100.0$ & $99.9$ & $100.2$ & $100.6$ & $101.0$ & $101.3$ & $101.8$ & $102.1$ & $102.3$ \\
& $0.6$ & $100.0$ & $96.4$ & $93.4$ & $91.4$ & $89.8$ & $88.5$ & $87.5$ & $86.6$ & $85.7$ \\
& $0.9$ & $100.0$ & $94.4$ & $88.5$ & $83.3$ & $79.8$ & $76.7$ & $73.9$ & $71.5$ & $69.6$ \\
& $0.95$ & $100.0$ & $99.2$ & $95.4$ & $90.9$ & $86.7$ & $83.1$ & $80.0$ & $77.3$ & $74.8$ \\
\noalign{\smallskip}\hline
\end{tabular}
\end{table}
Although the efficiencies are highly dependent across columns and are specifically based on model (\ref{Copula}), they still give some insight into the performance of pairwise likelihood estimators for asymptotically dependent distributions. Complementary results are provided by \citet[][p.148 and p.181]{Huser:2013} and \citet{Huser.Davison:2014a}. The efficiency of pairwise likelihood estimators decreases as $D$ increases but remains fairly high in moderate dimensions. For larger $D$, \citet{Huser:2013} suggests that the loss can be substantial. Moreover, the largest loss in efficiency seems to occur for $\alpha\approx 0.9$.

It is natural to wonder whether our results remain valid for other dependence structures, and future research is needed to explore asymmetric and non-Archimedean models. However, thus far it seems that estimator performance is most affected by the censoring scheme considered and overall dependence. In particular, for the asymmetric logistic model, or other models which put mass on the boundary faces or edges, non-censored methods cannot be used unless subtle adjustments are made.


\section{Discussion}
\label{DiscussionSection}
We have compared several likelihood estimators for the multivariate extreme-value logistic distribution. Our study shows that their performance is mainly influenced by the level of dependence, and by the ``weight'' attributed to each contribution to the likelihood function. Specifically, in moderate to weak dependence scenarios, threshold-based estimators tend to overestimate dependence, resulting in an overestimation of joint return levels. Non-censored estimators perform worst overall (except in strong dependence cases), but censored ones usually have a much better balance between bias and efficiency. The choice of the threshold is also crucial, since there is a trade-off between bias and variance. Our results suggest that higher thresholds should be considered when the dependence weakens. In high dimensions, where the bias is generally more pronounced, pairwise likelihood estimators behave best, because they are less sensitive to model misspecification. Interestingly, block maximum estimators also perform quite well in high dimensions when dependence is weak, but if the block size is constrained to be large, the smaller number of block maxima available results in higher variability, which might spoil the estimator. Although our results concern the logistic model, some preliminary investigations with the asymmetric logistic model suggest that the censored estimator work well more broadly. Further research is needed to explore cases in which a smoothness parameter must be estimated (but see \citealp{Thibaud.Opitz:2015}) and those where the dependence is strong between some variables but weak or nearly inexistent between others; such cases would be of particular interest for spatial applications. Finally, it would also be worth investigating cases where the speed of convergence to the limiting distribution is different to that used in our analysis. 

\appendix
\section*{Appendix: Asymptotic relative efficiencies}
\label{AsymptoticRelativeEfficienciesAppendix}
We detail below how the theoretical asymptotic relative efficiencies, reported in Table~\ref{AsymptoticRelativeEfficienciesTable}, are calculated. They are computed with the ratio of Fisher information quantities, i.e., assuming that block sizes and threshold probabilities are \emph{fixed}, whereas the sample size $n\to\infty$. Throughout, $(Y_1,Y_2)^T$ is supposed to be logistic distributed with unit Fr\'echet margins, i.e., $\pr(Y_1\leq y_1,Y_2\leq y_2)=\exp\{-V(y_1,y_2)\}$ with $V(y_1,y_2)=(y_1^{-1/\alpha}+y_2^{-1/\alpha})^\alpha$ for some $\alpha\in(0,1]$, while subscripts of the function $V$ denote partial differentiation with respect to the corresponding variables, e.g., $V_1=\partial V/\partial y_1$, $V_{12\alpha}=\partial^3 V/\partial y_1\partial y_2\partial \alpha$, etc. Similarly, the function $G=\exp(-V)$ denotes the logistic joint distribution, and $G_1=-V_1\exp(-V)$, $G_2=-V_2\exp(-V)$, $g=(V_1V_2-V_{12})\exp(-V)$ are its partial derivatives. The notation $\hat\alpha$ (with various subscripts) refers to the different estimators of $\alpha$.

\subsection*{A.1 Fisher information for block maximum estimators $\hat\alpha_{{\rm Max},1},\hat\alpha_{{\rm Max},2},\hat\alpha_{{\rm Max},3}$}
The Fisher information $i(\alpha)$ for the logistic model was derived by \citet{Shi:1995a}. For $n=LN$ independent observations and blocks of size $L$, the total Fisher information of $\hat\alpha_{{\rm Max},1}$ is $Ni(\alpha)$, and the average information per observation is $i_{{\rm Max},1}(\alpha)=Ni(\alpha)/n=i(\alpha)/L$. The Fisher information $i^\star(\alpha)$ for the logistic model when occurrence times of maxima are considered was derived by \citet{Stephenson.Tawn:2005}. Similarly, one obtains that the Fisher information per observation for $\hat\alpha_{{\rm Max},2}$ is $i_{{\rm Max},2}(\alpha)=i^\star(\alpha)/L$. For the bias-reduction approach \eqref{LoglikelihoodMax3} of \citet{Wadsworth:2015}, one can see that as the sample size $n$ and block size $L$ increases, the second-order likelihood term vanishes. This implies that the Fisher information $i_{{\rm Max},3}(\alpha)$ of $\hat\alpha_{{\rm Max},3}$ is approximately equal to $i_{{\rm Max},2}(\alpha)$ for large $L$. 

\subsection*{A.2 Fisher information for the threshold estimator $\hat\alpha_{{\rm Thr},4}$ and $\hat\alpha_{{\rm Thr},5}$ with marginal thresholds $\pmb{u}=(u,u)^T$}
For $\hat\alpha_{{\rm Thr},4}$, by definition of the censored contribution $p^1_u(y_1,y_2;\psi)$, the Fisher information of a single observation is
\begin{eqnarray}
i_{{\rm Thr},4}(\alpha)&=&i_{00}(\alpha) + i_{01}(\alpha) + i_{10}(\alpha) + i_{11}(\alpha)\label{decomp}\\
&=&\left\{-{\partial^2\over\partial\alpha^2}\log G(u,u)\right\} G(u,u) + \int_{u}^\infty\left\{-{\partial^2\over\partial\alpha^2}\log G_2(u,y_2)\right\}G_2(u,y_2){\rm d}y_2\nonumber\\
&&+\int_{u}^\infty\left\{-{\partial^2\over\partial\alpha^2}\log G_1(y_1,u)\right\}G_1(y_1,u){\rm d}y_1 + \int_{u}^\infty\int_{u}^\infty\left\{-{\partial^2\over\partial\alpha^2}\log g(y_1,y_2)\right\}g(y_1,y_2){\rm d}y_1{\rm d}y_2.\nonumber
\end{eqnarray}
By symmetry, one has $i_{10}(\alpha)=i_{01}(\alpha)$, and variants of Bartlett's identities then yield
\begin{eqnarray}
i_{00}(\alpha) &=& V_{\alpha^2}\exp(-V)\bigg|_{(u,u)},\nonumber\\
i_{10}(\alpha) &=& i_{01}(\alpha) = \left(V_{\alpha}^2-V_{\alpha^2}\right)\exp(-V)\bigg|_{(u,u)} + \int_{u}^\infty \left({V_{1\alpha}\over V_1}-V_{\alpha}\right)^2(-V_1)\exp(-V)\bigg|_{(u,y_2)}{\rm d}y_2,\label{i01}\\
i_{11}(\alpha) &=& -\left(V_{\alpha}^2-V_{\alpha^2}\right)\exp(-V)\bigg|_{(u,u)} \nonumber\\
&&+\int_{u}^\infty\int_{u}^\infty \left({V_{1\alpha}V_2+V_1V_{2\alpha}-V_{12\alpha}\over V_1V_2-V_{12}}-V_{\alpha}\right)^2(V_1V_2-V_{12})\exp(-V)\bigg|_{(y_1,y_2)}{\rm d}y_1{\rm d}y_2.\label{i11}
\end{eqnarray}
The integral in (\ref{i01}) can be transformed into a definite integral by the change of variable $v=V(u,y_2)$. After some calculations, one finds that this integral equals
\begin{equation}
\int_{u^{-1}}^{2^\alpha u^{-1}} {e^{-v}\over \alpha^2}\left[(1-v)v^{1/\alpha}(\log u+\log v) - \left\{1+\alpha(1-v)\left(v^{1/\alpha}-u^{-1/\alpha}\right)\right\}\log\left\{-1+(uv)^{1/\alpha}\right\}\right]^2{\rm d}v.\label{integr}
\end{equation}
A finite difference or standard Monte Carlo methods can then be used to compute (\ref{integr}) with high accuracy. The double integral in (\ref{i11}) can be markedly simplified by considering the same change of variables as for $\hat\alpha_{{\rm Thr},1}$, i.e., $v_1=V(y_1,y_2)$, $v_2=\{y_1 V(y_1,y_2)\}^{-1/\alpha}$. The program {\tt Mathematica} can then help in computing this integral analytically with respect to $v_2$, and a finite integration with compact support can be used to approximate the remaining complicated integral with respect to $v_1$.

To compute the Fisher information of the estimator $\hat\alpha_{{\rm Thr},5}$, minor changes may be applied to the decomposition in \eqref{decomp}, and calculations may then be done following the same lines. In particular, the same transformations of variables may be used to produce definite integrals that can be computed efficiently. In practice, if the threshold $u$ is large enough, then the tail approximations $\exp\{-V(y_1,y_2)\}$ and $1-V(y_1,y_2)$, $y_1,y_2>u$ are essentially similar (thanks to a first-order Taylor expansion of the exponential), and therefore the Fisher informations $i_{{\rm Thr},4}(\alpha)$ and $i_{{\rm Thr},5}(\alpha)$ are approximately equal for large $u$.

\begin{acknowledgements}
The authors thank Prof. Christian Genest for sharing his code to simulate from the Archimedean copula in dimension $D=2$, and Dr. Jennifer Wadsworth for fruitful discussions. Rapha\"el Huser was partly supported by the Swiss National Science Foundation.
\end{acknowledgements}

\bibliographystyle{spbasic}      
\bibliography{Biblio2}   


\end{document}